\documentclass[traditabstract]{aa} 
\usepackage{graphicx}
\usepackage{longtable}
\usepackage{txfonts}
\usepackage{natbib}
\bibpunct{(}{)}{;}{a}{}{,}
\begin{document}

\title{ {Stellar activity as a tracer of moving groups}\thanks{Full Table \ref{tblcatexample} will be only available in electronic form at the CDS.}}

   \author{F. Murgas
          \inst{1}
          \and
          J. S. Jenkins\inst{2}\fnmsep \inst{3}
          \and
          P. Rojo\inst{2}
          \and
          H.R.A Jones\inst{3}
           \and
           D.J. Pinfield\inst{3}   
          }

   \institute{Instituto de Astrof\'isica de Canarias (IAC), E-38205 La Laguna, Tenerife, Spain \\
   	        \email{murgas@iac.es}
          \and
          	 Departamento de Astronom\'ia, Universidad de Chile, Casilla Postal 36D, Santiago, Chile
          \and
          	 Center for Astrophysics Research, University of Hertfordshire, College Lane, Hatfield, Herts, UK, AL10 9AB\\
             }

   \date{Received 26 April, 2012; Accepted 13 February, 2013}

 
  \abstract
{We present the results of a study of the stellar activity in the solar neighborhood using complete kinematics (galactocentric velocities U,V,W) and the chromospheric activity index $\log R'_{\rm{HK}}$. We analyzed the average activity level near the centers of known moving groups using a sample of 2529 stars and found that the stars near these associations tend to be more active than field stars. This supports the hypothesis that these structures, or at least a significant part of them, are composed of kinematically bound, young stars. We confirmed our results by using Galaxy Evolution Explorer (GALEX) UV data and kinematics taken from the Geneva-Copenhagen Survey for the stars in the sample. Finally, we present a compiled catalog with kinematics and activities for 2529 stars and a list of potential moving group members selected based on their stellar activity level.}
{}{}{}{} 

   \maketitle


\section{Introduction}

Moving groups (MGs from now on) are groups of stars distributed across the sky that share the same velocity components and form streams. Unlike open clusters, they have no spatial or overdensity center discernible from the field stars. Since the stars that belong to these streams have the same motion, MGs are often called kinematic structures. The study of these structures has a long history (see \citealp{zuckerman&song_2004} for an introduction), but the modern era of MG studies began thanks mainly to the work of Olin Eggen (\citealp{Eggen58}). Nowadays, astronomers using data from large surveys like Hipparcos (\citealp{hipparcos}; \citealp{Hippnew}) have robustly confirmed the existence of MGs (\citealp{dehnen98}). The origin of MGs is still a question of some debate because a fairly large sample of stars is needed to confirm the presence of MGs and also because of the difficulty in identifying the membership of individual stars to these stellar structures.

The two main hypotheses that try to explain the origin of MGs are that either these structures are formed through dynamic interactions, or they have a cluster origin. The first theory postulates that the interaction between field stars and higher density zones (e.g., spiral arms, the galactic bar) produces an overdensity in velocity space, thus creating an MG. Evidence in favor of the dynamic or resonant origin of MGs is the success of computational models in reproducing some of the overdensities seen in velocity space (\citealp{dehnen2000}; \citealp{simone2004}; \citealp{antoja2009}; \citealp{bovy2010}, among others) and the differences in age (\citealp{famaey08}) of individual members that belong to these structures.

The second theory states that when the gas cloud that gives rise to a cluster or stellar association is cleared by winds and/or outflows driven by massive stars, some of the stars that belong to these structures are less affected by the gravitational force of the system. These stars could eventually become unbound through disk heating processes and leave their initial home. However, they conserve the velocity of the stellar system where they were born producing the streams observed in the sky that we called MGs. So far, the only MG that has been proven to have a cluster origin through an abundance study is HR 1614 (\citealp{desilva2007}), and at least a fraction of the stars that are members of the Hyades stream (\citealp{desilva_2011}, \citealp{Pompeia_2011}).

To shed some light onto this matter, we investigated the stellar activity level and kinematics of the stars in the sample of \citet{jenkins11} as well as the relationship between field stars and those towards the centers of known MGs.


\section{Activity estimators}
\subsection{Optical data}
For F, G, and K dwarfs a good indicator of activity is the Ca\textsc{ii} H and K lines at 3968.5 \AA{} and 3933.7 \AA{}, respectively. If the level of activity in the star is high, the H and K lines consist of a narrow emission component superimposed on the broad absorption line. The ratio between the chromospheric emission lines and the total bolometric emission of the star gives the activity index $R'_{HK}$ (\citealp{noyes1984}). 

In order to study these kinematic structures known as MGs, it is necessary to have the largest possible sample. With that objective in mind, we merged the Ca\textsc{ii} activity index catalogs of \citet{duncan1991}, \citet{henry1996}, \citet{gray2003}, \citet{gray2006}, and \citet{jenkins11}. In the case of \citet{duncan1991} catalog, we used the \citet{noyes1984} calibration to convert the measured S index to $\log R'_{\rm{HK}}$ values.

For the stars in common, \citet{jenkins08} found a discrepancy between their activity values and the ones computed by \citet{gray2006}. This can probably be explained by the use of different calibration stars between the samples. In \citet{jenkins11}, this offset was computed to be -0.15 dex, so we proceeded to add +0.15 dex to the $\log R'_{HK}$ obtained by \citet{gray2003} and \citet{gray2006}.

Jenkins et al. (2008, 2011) presents a spectroscopic study of F,G, and K dwarfs and subgiant stars in the southern hemisphere and computed radial, rotational ($v\sin i$), and galactocentric velocities (U,V,W) as well as the Ca\textsc{ii} activity index $\log R'_{HK}$ for stars in the sample. For our analysis we used 865 stars from \citet{jenkins11}. The other stellar activity catalogs do not have kinematic information and in order to obtain U,V,W velocities, a cross-match between the different catalogs and the Geneva-Copenhagen Survey (GCS, \citealp{GCSI}) was made. In the end, 1000 stars from the combined catalogs of \citet{gray2003} and \citet{gray2006}, as well as 790, 649, and 292 stars from \citet{henry1996}, \citet{wright2004}, and \citet{duncan1991}, respectively, had GCS velocities.

All the catalogs are unbiased in their selection criteria: \citet{duncan1991} measured everything observable with their instrument down to V= 12.4 mag; \citet{henry1996} selected G dwarfs based on their color; Gray et al. (\citeyear{gray2003} and \citeyear{gray2006}) selected dwarfs and giants earlier than M0 within 40pc of the Sun; and \citet{jenkins11} selected stars based on their color.

Leaving out the stars in common among the samples and removing 99 spectroscopic binaries found using SIMBAD\footnote{http://simbad.u-strasbg.fr/simbad/}, the final data set used in this analysis is composed of 2529 F, G, and K stars with complete stellar activity indices and kinematics. The compilation of catalogs with data for 2529 stars used in this paper will be available on-line for future use (Table \ref{tblcatexample}). The centers of known moving groups used in this work (Table \ref{tblcenters}) were taken from \citet{antoja2009}, except Castor which was taken from \citet{montes2001}.

\begin{table}
      \caption{First ten lines of the full sample of 2529 stars. This catalogue will be available on line.}
	       $$ 
         \begin{array}{p{0.5\linewidth}crrrr}
            \hline
            \noalign{\smallskip}
              Star & U (km/s) & V (km/s)  & W (km/s) & \log R'_{\rm{HK}} \\
            \noalign{\smallskip}
            \hline
            \noalign{\smallskip}
			HD 23   &   40.00 & -22.00 & -16.00 & -4.780^{(3)} \\
			HIP 72   & -48.95 & -98.00 &   19.87 & -5.090^{(1)} \\
			HD 105 & -10.00 & -21.00 &   -1.00 & -4.360^{(3)} \\
			HD 142 & -58.00 & -38.00 & -16.00 & -4.703^{(2)} \\
			HD 166 & -15.00 & -22.00 & -10.00 & -4.308^{(2)} \\
			HD 361 &     3.00 &     2.00 &   -7.00 & -4.684^{(2)} \\
			HD 377 & -14.00 &   -7.00 &   -4.00 & -4.360^{(4)} \\
			HD 400 &   28.00 & -10.00 &   -8.00 & -4.870^{(4)} \\
			HD 564 &   26.00 &     4.00 &   -4.00 & -4.830^{(3)} \\
			HIP 599 &   15.89 &   11.16 &   -7.93 & -4.990^{(1)} \\
            \noalign{\smallskip}
            \hline
         \end{array}
     $$ 
     \label{tblcatexample}
   \tablefoot{Activities from: (1) \citet{jenkins11}, (2) Gray et al. (\citeyear{gray2003}, \citeyear{gray2006}), (3) \citet{henry1996}, (4) \citet{wright2004}, (5) \citet{duncan1991}.}  
   \end{table}

Of the final sample of 2529 stars, 2502 have a B-V color index measured by Hipparcos. Using this information, 632 stars can be classified as F stars ($(B-V) < 0.58$), 1577 as G stars ($0.58 < (B-V) < 0.81$), and 293 as K stars ($(B-V) > 0.81$). Hence, G stars are the dominant spectral type in the sample. From the catalog of \citet{lopezsantiago_2009}, 153 stars of our sample are classified as MG members; and from \citet{Torres_2006} 134 stars share the same status.

\begin{table}
      \caption{Centers of known moving groups taken from \citet{antoja2009} and \citet{montes2001}.}
	       $$ 
         \begin{array}{p{0.5\linewidth}crr}
            \hline
            \noalign{\smallskip}
              Group name & U (km/s) & V (km/s) \\
            \noalign{\smallskip}
            \hline
            \noalign{\smallskip}
            Pleiades &  -12 &  -22 \\
			Hyades  &  -40 &  -20  \\
			Sirius/UMa & 9 & 3 \\
			Coma Berenices & -10 & -5 \\
			NGC 1901 & -25 & -10 \\
			HR 1614 & 15 & -60 \\
			Dehnen 7   & 20 & -20 \\
			Dehnen 8 & -40 & -50 \\
			Dehnen 9 & -25 & 50 \\
			Dehnen 10 & 50 & 0 \\
			Dehnen 11  & 50 & -25 \\
			IC 2391 & -20.8 & -15.9 \\
			Dehnen 13 & -70 & -10 \\
			Dehnen 14 & -70 & 50 \\
			61 Cygni & -80 & -53 \\
			$\zeta$ Herculis & -30 & -50 \\
			Wolf 630 & 25.8 & -33 \\
			Castor & -10.7 & -8.0 \\
            \noalign{\smallskip}
            \hline
         \end{array}
     $$ 
     \label{tblcenters}
   \end{table}
   
\begin{figure}
   \centering
  	\includegraphics[width=\linewidth]{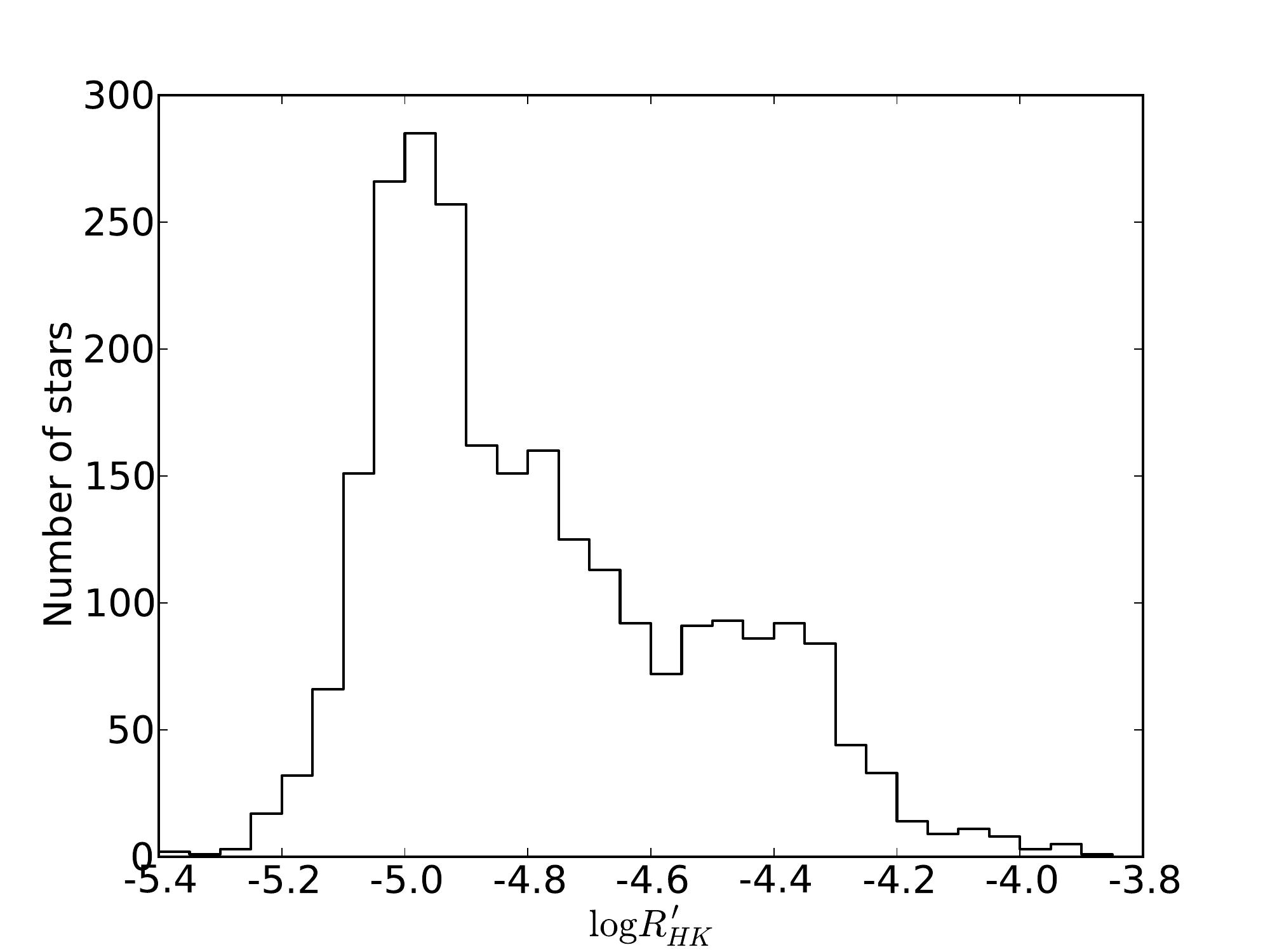}
      \caption{Chromospheric activity index distribution for the entire sample.}
      \label{logRdistri}
   \end{figure}

Figure \ref{logRdistri} shows the distribution of chromospheric activity for the entire sample. The distribution presents a bimodal feature with peaks representing active stars at $\sim -4.4$ dex and inactive stars at $\sim -5.0$ dex, in agreement with previous results (\citealp{duncan1991} \citealp{henry1996}; \citealp{gray2006}; \citealp{jenkins06}). The region in the $\log R'_{\rm{HK}}$ distribution for solar-type dwarfs and subgiants that is located between the bimodal active and inactive peaks in this distribution is known as the Vaughan-Preston (VP) gap. Older and inactive stars are located beyond this gap with much lower activity and so can serve as a good proxy for younger stars. For descriptions of the VP gap region we refer the reader to other works (e.g., \citealp{vaughan80}; \citealp{baliunas95}; \citealp{henry1996}; \citealp{jenkins11}). 
 
\subsection{UV data}
\begin{figure}
	\centering
	\includegraphics[width=\linewidth]{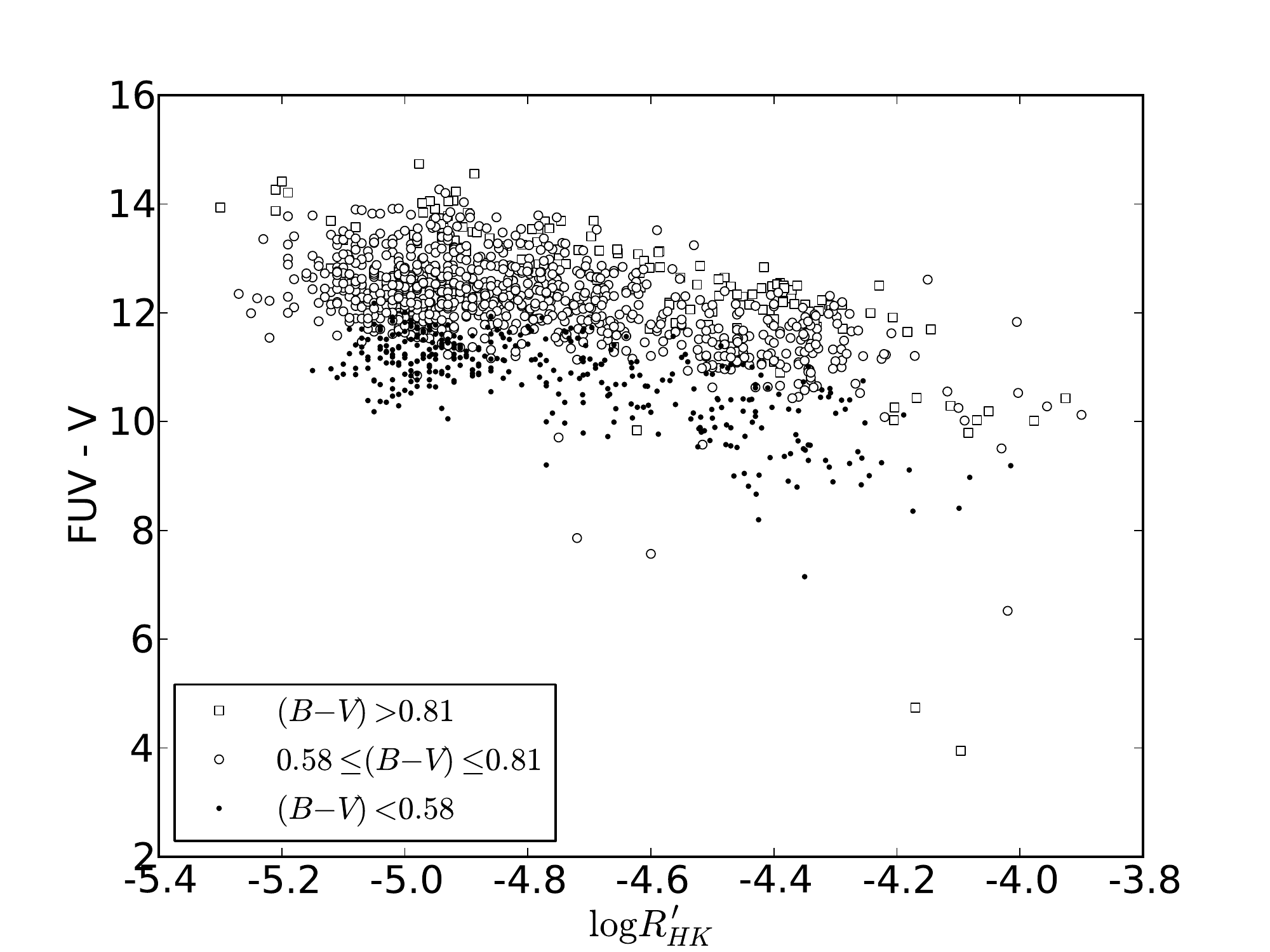}
	\caption{FUV - V versus $\log R'_{\rm{HK}}$ for a sub-sample of 1360 stars.}
	\label{FUVvsRhk}
\end{figure}

Since the UV flux has been shown to be correlated with chromospheric activity, as an extra test of our results we obtained UV photometry using the GALEX All-Sky Survey (\citealp{galex}) for the entire data set. GALEX ultraviolet bands, far-UV (FUV, 1350 - 1780\AA{}) and near-UV (NUV, 1780 - 2830\AA{}), were acquired by doing a search for the nearest source around the Hipparcos coordinates with a search radius of 10 arcsec. To have a distance-independent indicator, we computed the FUV - V color using Hipparcos photometry. \citet{findeisen11} have shown that the color FUV - V  is roughly correlated with the calcium HK index (see Fig. \ref{FUVvsRhk}) and age. From our total sample of 2529 stars, only 1360 had FUV and V magnitudes.

\begin{figure*}
	\includegraphics[width=\linewidth]{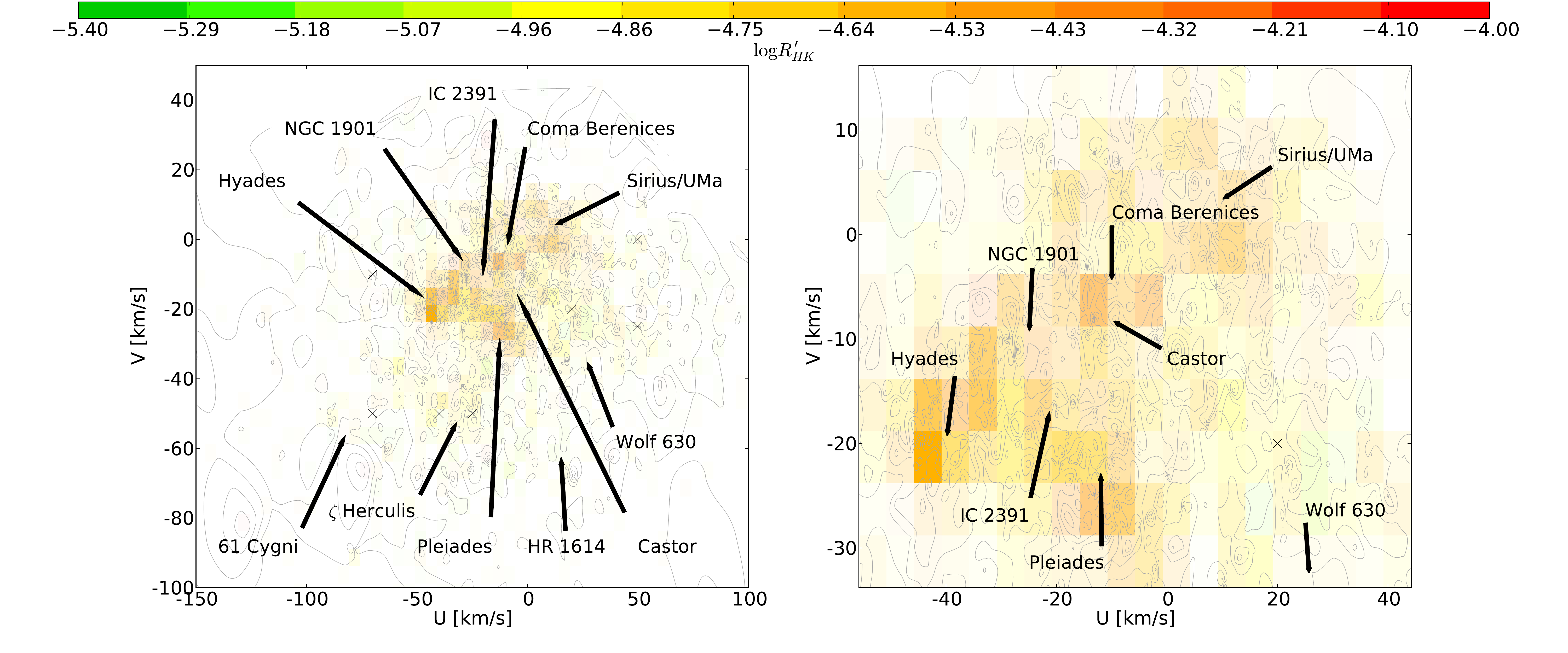}
	\caption{Plot of the (U,V) velocity space showing contour levels of stellar activity ($\log R'_{\rm{HK}}$) and the grid of boxes for the sample (left). The right panel shows a zoom of the central region where a higher density of stars is found. The transparency level indicates the density number of stars inside the boxes. The crosses mark the center of the Dehnen 7, 8, 9, 10, 11, 13, 14 groups.}
	\label{contour1}
\end{figure*}

\section{Results}
\subsection{Data analysis}
We plotted the contour levels of the U,V,$\log R'_{\rm{HK}}$ distribution for the data set in order to detect the structures that presented higher levels of stellar activity in kinematic velocity space. To confirm the detection of active regions we created a grid of boxes of 5 km/s in size in U,V ($\sim$2$\sigma$ precision in space velocity of \citealp{jenkins11}) and proceeded to compute the average activity level inside the box. Figure \ref{contour1} shows the activity index contours versus velocity space for the entire sample (left) and a zoom of the region with the highest density of stars (right); the level of transparency indicates the density number inside each box which is normalized to the box with the highest number of stars in the data set. From the figure we can see that some centers of known moving groups are near regions with active stars.

In Table \ref{tblactivboxs} we present the mean and standard deviation of the stellar activity inside boxes that contain the center of MGs with more than three stars. According to our grid, one box contained the centers of two MGs, Coma Berenices and Castor, which is not surprising given their proximity and the resolution of our grid.

Six of the groups analyzed here present an average activity level between -4.75 and -4.1 dex which is considered to be the active subset (\citealp{henry1996}); the rest are considered inactive. The active MGs are consistent with the cluster related origin, since these are also the youngest MGs (\citealp{montes2001}). The inactive groups (Dehnen 7, 9, 10, 11, and $\zeta$ Herculis) present a $\log R'_{\rm{HK}}$ in the range $-4.78$ - $-4.86$ and according to the activity-age relation of \citet{mamajek&hillenbrand2008}, these MGs have ages between 3.0 and 4.1 Gyr.
   
   \begin{table}
      \caption{Average stellar activity of MGs inside boxes with more than three stars.}
	       $$ 
         \begin{array}{p{0.5\linewidth}crrr}
            \hline
            \noalign{\smallskip}
              Group name & \log R'_{\rm{HK}} & \sigma_{dev} & N_{stars} \\
            \noalign{\smallskip}
            \hline
            \noalign{\smallskip}
				
				Pleiades  & -4.66 & 0.27 & 24 \\
				Hyades  & -4.66 & 0.24 & 25 \\
				Sirius/UMa  & -4.61 & 0.27 & 9 \\
				Coma Berenices + Castor  & -4.59 & 0.25 & 15 \\
				NGC 1901  & -4.53 & 0.16 & 11 \\
				IC 2391  & -4.58 & 0.26 & 21 \\
				Dehnen 7 & -4.81 & 0.25 & 9 \\
				Dehnen 9 & -4.86 & 0.23 & 6 \\
				Dehnen 10 & -4.78 & 0.19 & 5 \\
				Dehnen 11 & -4.86 & 0.18 & 4 \\
				$\zeta$ Herculis  & -4.87 & 0.22 & 6 \\
				
            \noalign{\smallskip}
            \hline
         \end{array}
     $$ 
     \label{tblactivboxs}
   \end{table}
   
\begin{figure}
    \centering
  	\includegraphics[width=\linewidth]{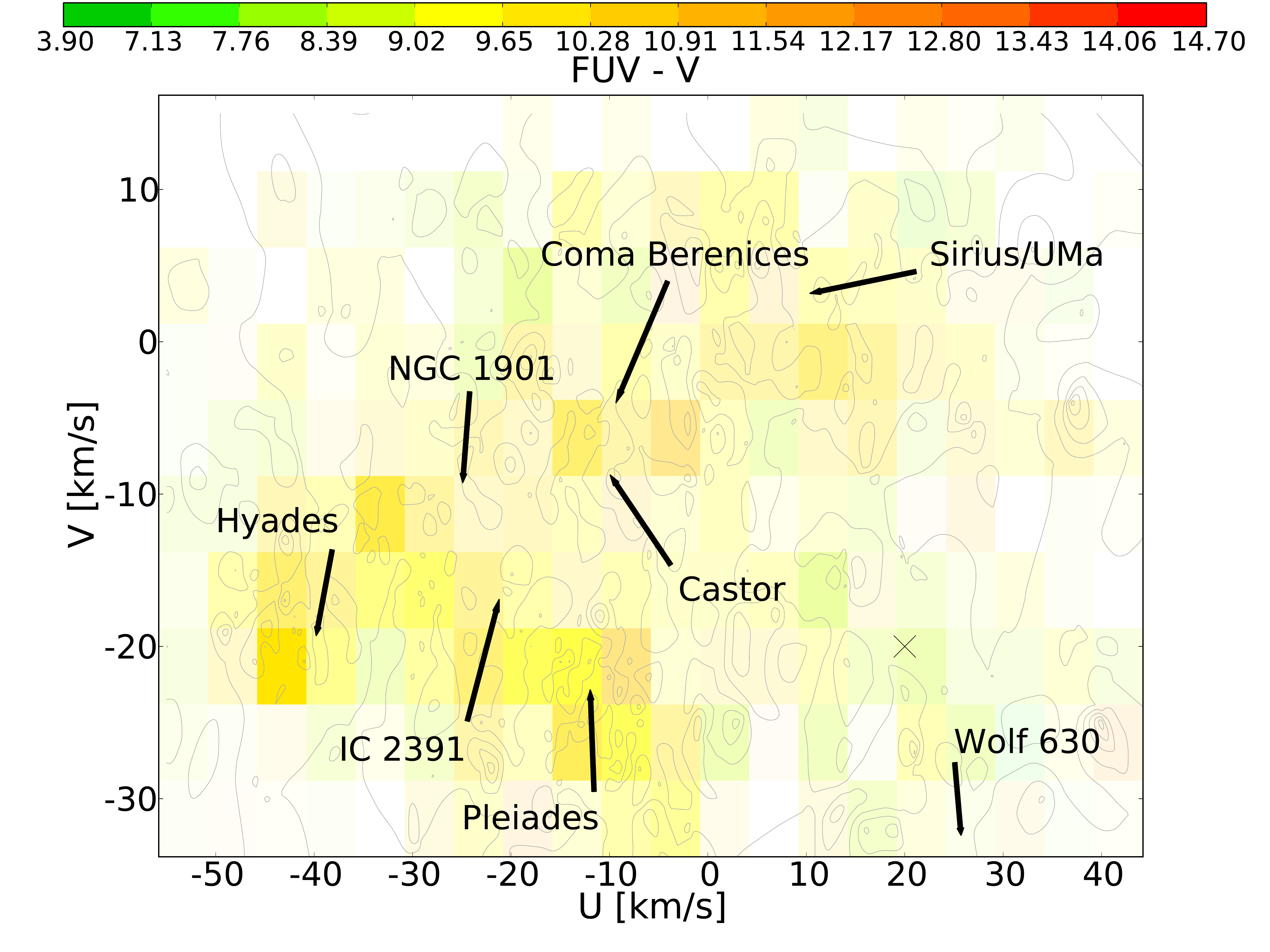}
    \caption{Plot of the (U,V) region where a higher density of stars is found. Contour levels and grid of the FUV-V for a sample of 1360 stars are shown.}
    \label{contour3}
\end{figure}

The same analysis was done on the FUV-V color, and the result (Fig. \ref{contour3}) is very similar to the one found with the activity index, which was expected since the FUV-V is correlated with $\log R'_{\rm{HK}}$.  

\subsection{Statistical tests}

In order to test the robustness of our claim that areas around MGs in kinematic velocity space generally host more active stars, we ran some bootstrap simulations to generate random data sets drawn from the observed data to test how often the MG centers would appear to host inactive boxes at our grid resolution. This allowed us to test how often randomness would give rise to these results in our data. To do this, we randomly scrambled the mean activities of our boxes to generate fake data sets, but crucially, only used boxes that contain at least three stars, which is necessary to ensure as best we can that we are not biasing our data through small random fluctuations because of low number statistics.  We ran the bootstrap simulation 500000 times to ensure a high level of statistical significance in our final probabilities.

We decided to test our data set in three different ways: a) the fraction of MG boxes that are more active than the VP region ($\log R'_{\rm{HK}} > -4.70$) versus the inactive boxes, b) the ratio between this first MG fraction (MG$_{VP}$) and the same fraction for field boxes, and c) the ratio between the mean activity of boxes in MGs compared with the same mean activity for boxes in the field.

\begin{figure}
   \centering

  	\includegraphics[width=8.0 cm]{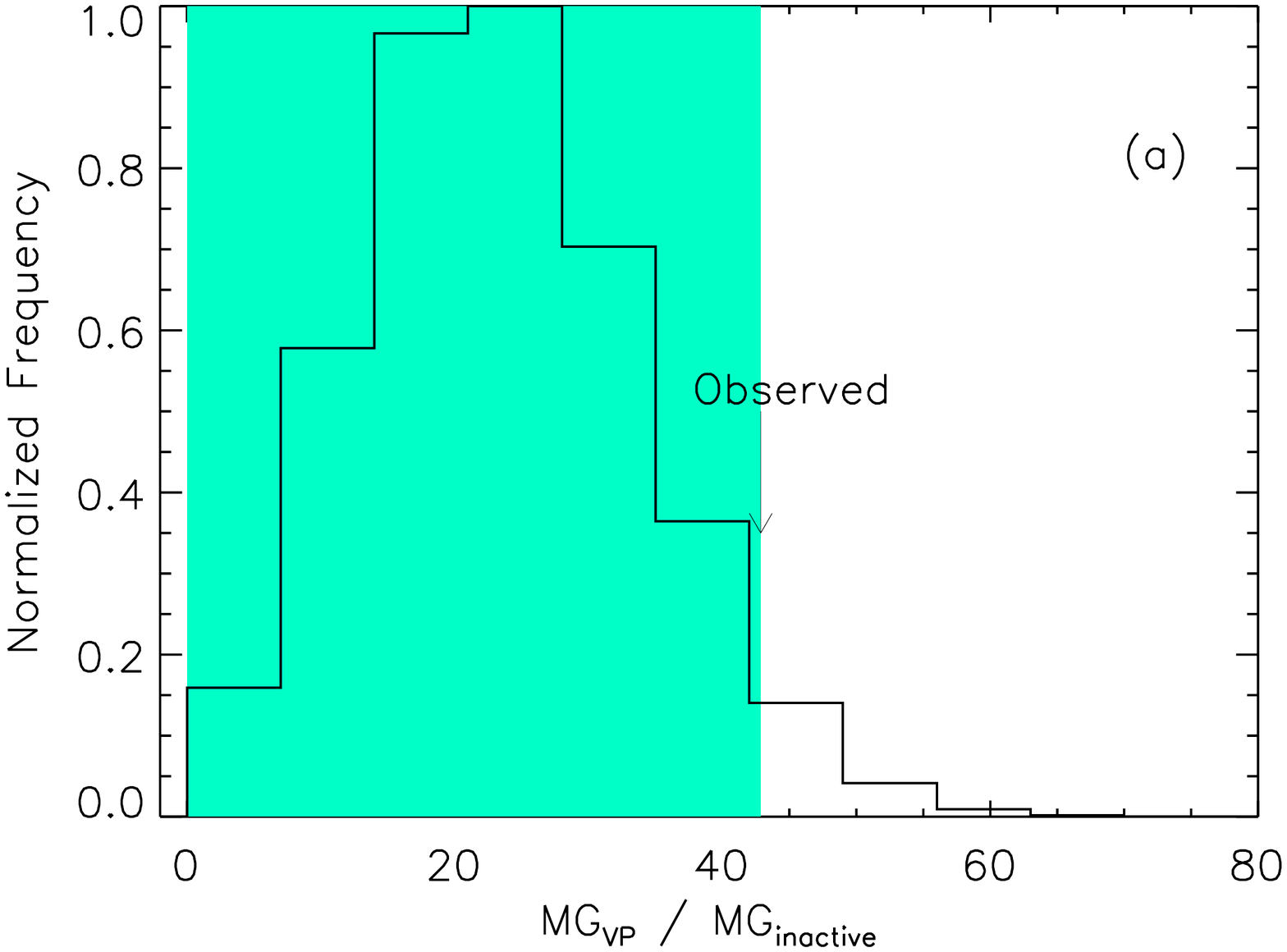}
  	\includegraphics[width=8.0 cm]{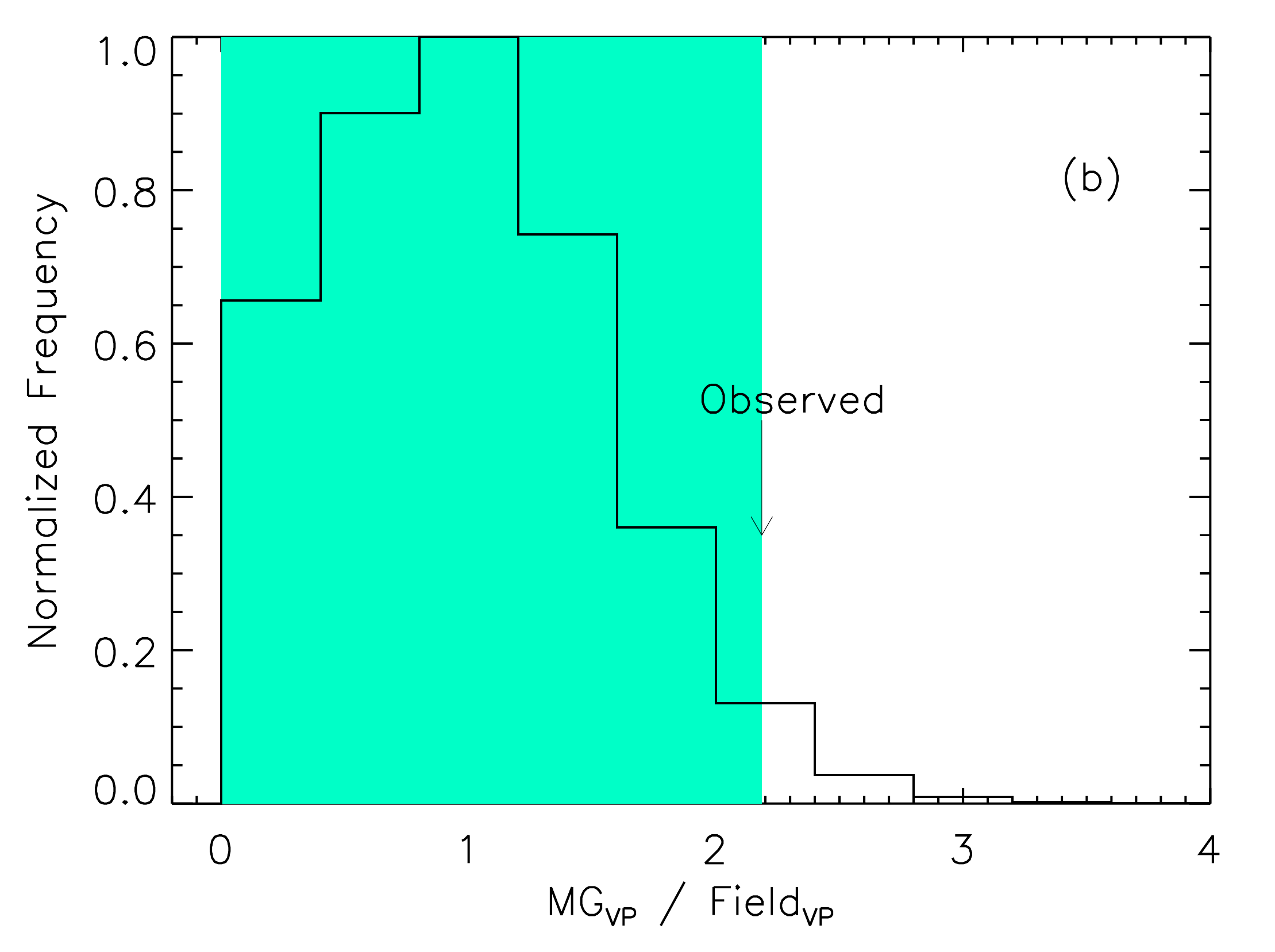}
  	\includegraphics[width=8.0 cm]{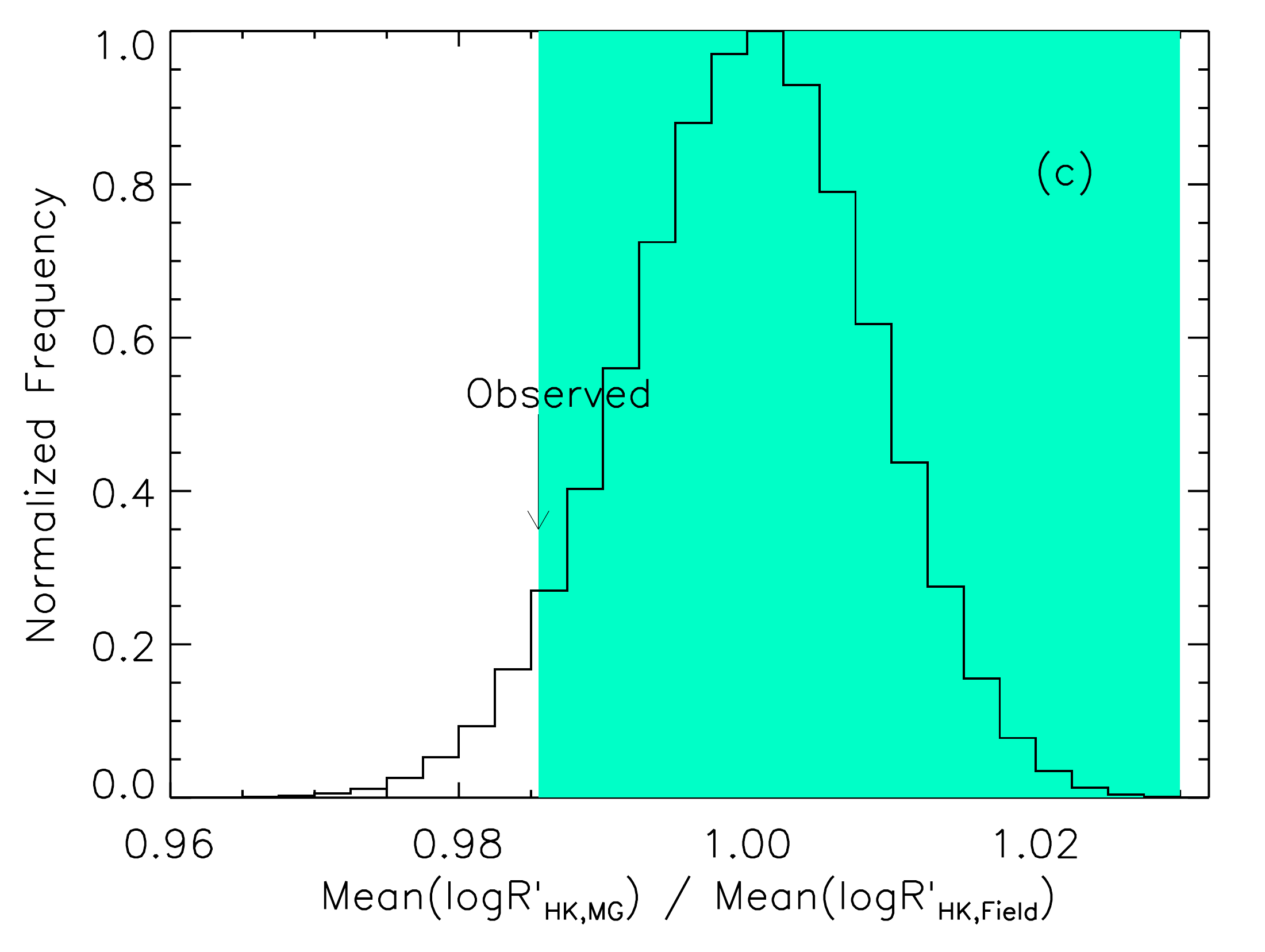}

      \caption{Bootstrap simulation tests of our data set: a) test the fraction of MG boxes that are more active than the VP region, b) the ratio between this first MG fraction and the same fraction for field boxes and c) the ratio between the mean activity of boxes in MGs compared with the same mean activity for boxes in the field. The original observed values are highlighted by the downward pointing arrows and the green shaded areas represent the percentage of times where the MG boxes showed a higher level of activity for each test.}
      \label{stattests}
\end{figure}

The bootstrap results for the first test are shown in Fig. \ref{stattests}a where the normalized frequencies of these ratios are shown and the original observed value is highlighted by the downward pointing arrow. Only 4.58\% of scrambled runs were found to host a set of MG boxes with a higher ratio of more active values than the original data set (unshaded area in Fig. \ref{stattests}a).

The second test (Fig. \ref{stattests}b), where we compared the fraction of boxes that are more active than the VP region against the field, shows a distribution that appears to be Poissonic. This is not surprising, given that this is essentially based on counting statistics. Again the original observed value is shown by the downward pointing arrow and, as mentioned, we find a similar value of only 4.58\% for ratios with more active field stars beyond the VP gap compared to the MGs, meaning we have 95.4\% confidence in our results.

The final test we decided to implement was to then compare the overall mean values between the MG boxes and the field boxes in our sample. This provided a robust test across the entire sample and increased the resolution of our bootstrap method given that we were sampling a larger number within our parameter-space.  Figure \ref{stattests}c shows the results from this bootstrap analysis, and now we clearly see a Gaussian shape and much better resolution than in the first two tests. Again, the observed value is shown by the downward pointing arrow. This time we find only 4.8\% of runs have a higher mean activity in the field compared to the mean activity of the MG centers, a little more than a $2\sigma$ result. By focusing on only the youngest MG, and those with the most stars in our grid we could increase the significance of this result, but this would introduce a bias into our methodology, so we have decided to include all known MGs for the time being.
  
\begin{figure}
      \centering
	  \includegraphics[width=\linewidth]{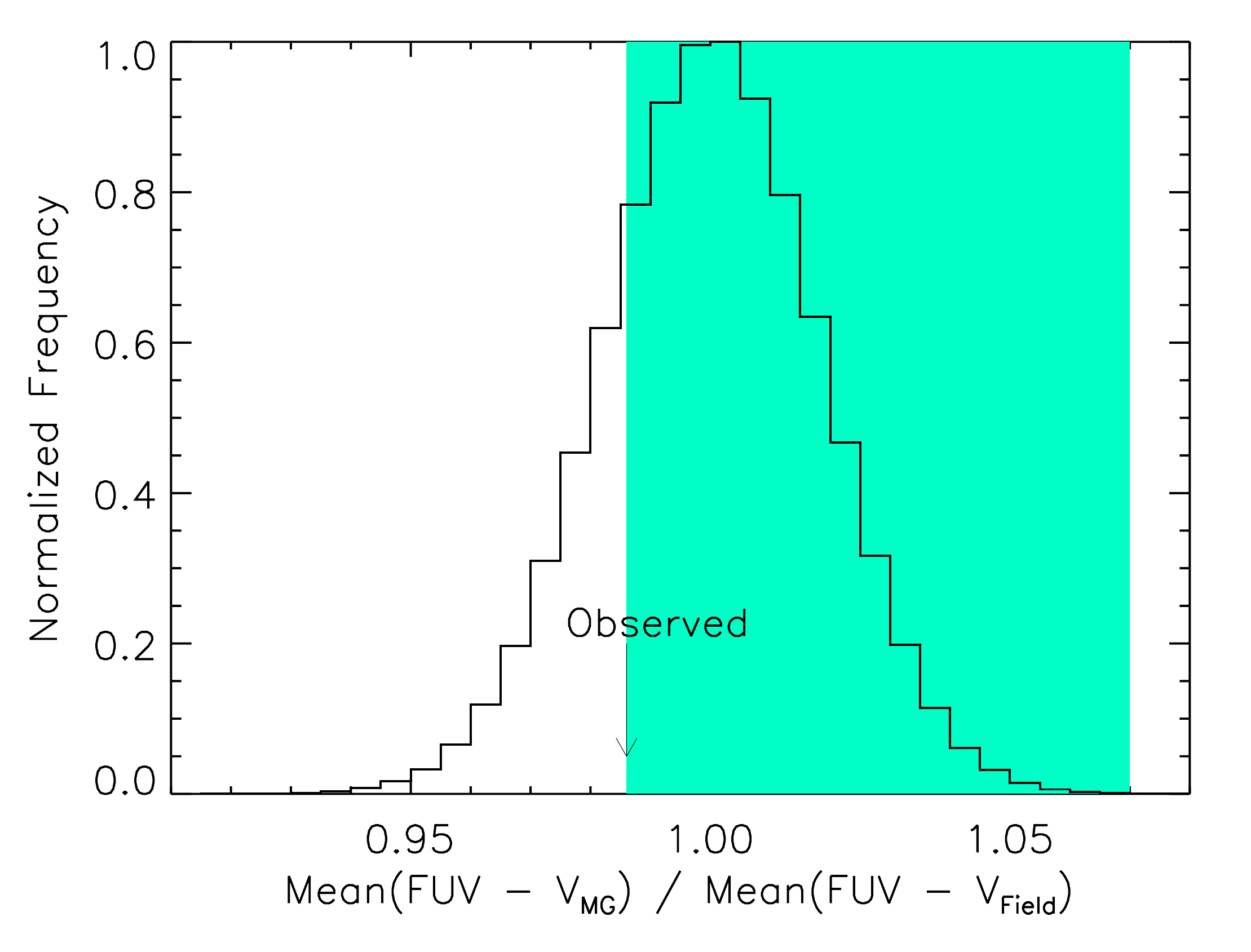}
      \caption{Similar to Fig. \ref{stattests}c, bootstrap simulation test showing the ratio between the mean FUV - V color of boxes that host MGs compared with the boxes in the field.}
      \label{stattests2}
\end{figure}

The same bootstrap analysis was performed using the FUV - V color as an activity indicator for the MG boxes (Fig. \ref{stattests2}) and we found that in 23.89\% of the runs the field stars have a higher stellar activity than the MG. The UV data confirm our results but not at the same level of significance as the $\log R'_{\rm{HK}}$ index. This is explained because the FUV - V color is only weakly correlated with the calcium index and the relation between these two quantities is not strong; the purpose of using UV data was to have an independent activity indicator of the results found with the $\log R'_{\rm{HK}}$ index using a large and homogeneous survey like GALEX.
   
Since stellar chromospheric activity depends on the mass of the star, we performed the same statistical tests separating the stars in the sample by their spectral classification using their (B-V) color index. For F and K stars, the results for the tests are similar: Test a) 74.7\% and 64.1\% of the time we found boxes with higher activities than the original set; Test b) in 38.5\% and 33.3\% of the runs, field stars presented a higher activity level than theMGboxes; and Test c) 62.5\% and 49.1\% of the time the mean activity level of the field surpassed the level of MG boxes. This result contradicts the statistics found by using the full sample (without separating the stars by spectral classification), but we believe this was caused by the low number of F and K stars in our sample (25.3\% and 11.7\% of the total, respectively). This low number of F and K stars is even more evident when we look at the most populated boxes that contain a known MG, i.e., the Hyades, Pleiades, and IC 2391. Of a total of 25 stars in the box containing the Hyades, only 8 and 2 stars are F and K types; for Pleiades, of 24 stars only 4 and 6 are F and K, respectively; and finally, for IC 2391 only 7 and 1 stars have (B-V) color indices consistent with F and K spectral types. The low number of F and K stars in our sample affects the estimation of the stars that are more active than the VP gap for test a) and the level of activity of the field for tests b) and c), in the sense that there are fewer boxes with more than three stars to compute the mean level of $\log R'_{\rm{HK}}$. This produces contradictory results when compared with the full sample and with the results found using only G stars.

For G stars, the results for the different tests are: 24.1\%, 6.2\%, and 15.9\% for tests a), b), and c), respectively. This is more in line with the results found by using the full sample, which was expected since the sample is dominated by G stars (63\% of the total).

Given these three tests made with $\log R'_{\rm{HK}}$ for the entire sample, the one performed with the FUV - V color index, and separating the stars by spectral type, we find that MG regions in kinematic space are generally more active than in the field at around the 2$\sigma$ level, reinforcing the hypothesis that, in general, most of these kinematic regions are actual overdensities of young stars and so are true MGs and not dynamic resonances attributed to structures in the galactic disk of our galaxy.


\subsection{Candidates}
\addtocounter{table}{1}

In Table \ref{tblcandidates} we present a list of active stars ($\log R'_{\rm{HK}} \geq -4.70$) that belong to the boxes in the grid that contained the center of a known MG. Of our list of candidates, HD 206860, HD 142415, HD 13507, HD 40979, and HD 73256 have known low mass companions or planets. Some of the stars presented here are classified as possible members of MGs by previous studies (e.g., \citealp{montes2001}, \citealp{lopezsantiago_2006}, \citealp{Maldonado_2010}, \citealp{Nakajima_2012}). A total of 42 stars are not cataloged and they could be new MG members.


\section{Discussion}
Based on our data sample, the statistical tests favor the idea that the MGs have a higher fraction of active stars than the field. Given our current theories of stellar formation and evolution, it is expected that young stars present a higher activity level when compared with older stars. This implies that at least a significant part of the stars that belong to these structures must be relatively young and since they are occupying the same velocity space, it is highly likely that they have the same origin.

The mean survival time of an MG is $\sim 1$ Gyr before being dispersed by disk heating mechanisms (\citealp{desimone04}); according to the activity-age relation from \citet{mamajek&hillenbrand2008} a star with an age of 1 Gyr should have a calcium activity index of approximately $-4.56$. From our results, the MGs that have an age similar to 1 Gyr are Pleiades, Hyades, Sirius/UMa, Coma Berenices, Castor, NGC 1901, and IC 2391. All of these MGs are known to be young, with ages less than 1 Gyr (\citealp{montes2001}), so a higher chromospheric activity level was expected. The groups Dehnen 7, 9, 10, 11, and $\zeta$ Herculis present ages ranging from 3 to 4.1 Gyr, but even if they are older than the canonical disruption time scale, according to our results they also have a higher stellar activity level when compared to the field stars that are near the center of the MG. In other words, the old MGs also conserve a signature showing that those stars are somehow related because they have an activity level higher than their environment and share the same velocity.

In the case that these structures are indeed formed by dynamic or resonant interactions, one would expect that the stars that composed the system would have different origins and hence different stellar parameters such as metallicity. Several studies have shown that the stars that belong to MGs have a wide spread of chemical compositions as opposed to an open cluster (e.g., \citealp{Bensby_2007}; \citealp{antoja2009}), but this could be caused by contamination in the selection of stars. 

Given our results, we think that the calcium activity index could be used as a parameter to pre-select stars as candidate MG members, which would then be confirmed through an abundance study. We plan to study the abundances of the active MG stars from the \citet{jenkins11} sample; metallicities are currently being processed for the entire data set. However, we note that generally high activity means high rotation and therefore a higher fraction of blended lines in stellar spectra, which leads to less robust atomic abundance measurements.


\section{Conclusions and future work}
We have presented a stellar activity study of moving groups using a sample of 2529 stars. By creating a grid of boxes in (U,V) velocity space and taking the mean activity inside the boxes, we find that, on average, the moving groups present a higher level of stellar activity than the field stars do. We find a similar result, although with a less statistical significance, by repeating the procedure with ultraviolet data taken from GALEX and for the G stars in the sample. Finally, a list of possible MG members with a high level of stellar activity is presented. We speculate that if this higher level of chromospheric activity in MG is a consequence of the fact that these structures are mainly composed of young stars, this is strong evidence that stars that belong to these streams form a real physical structure, that they have a common origin, and that they are not structures formed by resonance interactions.

We plan to follow-up these MG members, and those in the literature, with high-contrast adaptive optics (AO) imaging techniques to hunt for orbiting sub-stellar companions. In particular, we shall focus on the lowest-mass MG members, the later M-types (e.g., \citealp{Clarke2010}; \citealp{Galvez2010}), where current AO imaging can reach down into the Jupiter-mass regime.  Late-type M-dwarfs are soon to provide large increase in the number of low-mass planets because methods and techniques are being pushed to hunt for such planets using the radial velocity and transit method (\citealp{nutzman_2008}; \citealp{ramsey_2008}; \citealp{jenkins09}; \citealp{Barnes2012}). However, such techniques are biased toward the oldest, most quiescent stars, and therefore direct imaging of young MG members can help to extend the parameter space in the search for planetary-mass objects around the lowest-mass stars.


\begin{acknowledgements}
F. Murgas is supported by RoPACS, a Marie Curie Initial Training Network funded by the European Commission’s Seventh Framework Programme. J.S. Jenkins acknowledges funding by Fondecyt through grant 3110004 and partial support from Centro de Astrof\'\i sica FONDAP 15010003, the GEMINI-CONICYT FUND, and from the Comit\'e Mixto ESO-GOBIERNO DE CHILE. P. Rojo also acknowledges support from Fondecyt 11080271, 1120299, Basal Center for Astrophysics and Applied Technologies, and FONDAP for astrophysics. This research has made use of the SIMBAD database, operated at CDS, Strasbourg, France. Some of the data presented in this paper were obtained from the Multimission Archive at the Space Telescope Science Institute (MAST). STScI is operated by the Association of Universities for Research in Astronomy, Inc., under NASA contract NAS5-26555. Support for MAST for non-HST data is provided by the NASA Office of Space Science via grant NNX09AF08G and by other grants and contracts.

\end{acknowledgements}


	\bibliographystyle{aa}
	\bibliography{biblio.bib}

\begin{thebibliography}{47}
\expandafter\ifx\csname natexlab\endcsname\relax\def\natexlab#1{#1}\fi

\bibitem[{{Antoja} {et~al.}(2008){Antoja}, {Figueras}, {Fern{\'a}ndez}, \&
  {Torra}}]{antoja2009}
{Antoja}, T., {Figueras}, F., {Fern{\'a}ndez}, D., \& {Torra}, J. 2008, \aap,
  490, 135

\bibitem[{{Baliunas} {et~al.}(1995){Baliunas}, {Donahue}, {Soon}, {Gilliland},
  \& {Soderblom}}]{baliunas95}
{Baliunas}, S.~L., {Donahue}, R.~A., {Soon}, W., {Gilliland}, R., \&
  {Soderblom}, D.~R. 1995, in Bulletin of the American Astronomical Society,
  Vol.~27, American Astronomical Society Meeting Abstracts \#186, 839--+

\bibitem[{{Barnes} {et~al.}(2012){Barnes}, {Jenkins}, {Jones}, {Rojo},
  {Arriagada}, {Jord{\'a}n}, {Minniti}, {Tuomi}, {Jeffers}, \&
  {Pinfield}}]{Barnes2012}
{Barnes}, J.~R., {Jenkins}, J.~S., {Jones}, H.~R.~A., {et~al.} 2012, \mnras,
  424, 591

\bibitem[{{Bensby} {et~al.}(2007){Bensby}, {Oey}, {Feltzing}, \&
  {Gustafsson}}]{Bensby_2007}
{Bensby}, T., {Oey}, M.~S., {Feltzing}, S., \& {Gustafsson}, B. 2007, \apjl,
  655, L89

\bibitem[{{Boesgaard} \& {Budge}(1988)}]{boesgaard1988}
{Boesgaard}, A.~M. \& {Budge}, K.~G. 1988, \apj, 332, 410

\bibitem[{{Bovy} \& {Hogg}(2010)}]{bovy2010}
{Bovy}, J. \& {Hogg}, D.~W. 2010, \apj, 717, 617

\bibitem[{{Clarke} {et~al.}(2010){Clarke}, {Pinfield}, {G{\'a}lvez-Ortiz},
  {Jenkins}, {Burningham}, {Deacon}, {Jones}, {Pokorny}, {Barnes}, \&
  {Day-Jones}}]{Clarke2010}
{Clarke}, J.~R.~A., {Pinfield}, D.~J., {G{\'a}lvez-Ortiz}, M.~C., {et~al.}
  2010, \mnras, 402, 575

\bibitem[{{De Silva} {et~al.}(2007){De Silva}, {Freeman}, {Bland-Hawthorn},
  {Asplund}, \& {Bessell}}]{desilva2007}
{De Silva}, G.~M., {Freeman}, K.~C., {Bland-Hawthorn}, J., {Asplund}, M., \&
  {Bessell}, M.~S. 2007, \aj, 133, 694

\bibitem[{{de Silva} {et~al.}(2011){de Silva}, {Freeman}, {Bland-Hawthorn},
  {Asplund}, {Williams}, \& {Holmberg}}]{desilva_2011}
{de Silva}, G.~M., {Freeman}, K.~C., {Bland-Hawthorn}, J., {et~al.} 2011,
  \mnras, 415, 563

\bibitem[{{De Simone} {et~al.}(2004{\natexlab{a}}){De Simone}, {Wu}, \&
  {Tremaine}}]{simone2004}
{De Simone}, R., {Wu}, X., \& {Tremaine}, S. 2004{\natexlab{a}}, \mnras, 350,
  627

\bibitem[{{De Simone} {et~al.}(2004{\natexlab{b}}){De Simone}, {Wu}, \&
  {Tremaine}}]{desimone04}
{De Simone}, R., {Wu}, X., \& {Tremaine}, S. 2004{\natexlab{b}}, \mnras, 350,
  627

\bibitem[{{Dehnen}(1998)}]{dehnen98}
{Dehnen}, W. 1998, \aj, 115, 2384

\bibitem[{{Dehnen}(2000)}]{dehnen2000}
{Dehnen}, W. 2000, \aj, 119, 800

\bibitem[{{Duncan} {et~al.}(1991){Duncan}, {Vaughan}, {Wilson}, {Preston},
  {Frazer}, {Lanning}, {Misch}, {Mueller}, {Soyumer}, {Woodard}, {Baliunas},
  {Noyes}, {Hartmann}, {Porter}, {Zwaan}, {Middelkoop}, {Rutten}, \&
  {Mihalas}}]{duncan1991}
{Duncan}, D.~K., {Vaughan}, A.~H., {Wilson}, O.~C., {et~al.} 1991, \apjs, 76,
  383

\bibitem[{{Eggen}(1958)}]{Eggen58}
{Eggen}, O.~J. 1958, \mnras, 118, 65

\bibitem[{{Famaey} {et~al.}(2008){Famaey}, {Siebert}, \& {Jorissen}}]{famaey08}
{Famaey}, B., {Siebert}, A., \& {Jorissen}, A. 2008, \aap, 483, 453

\bibitem[{{Findeisen} {et~al.}(2011){Findeisen}, {Hillenbrand}, \&
  {Soderblom}}]{findeisen11}
{Findeisen}, K., {Hillenbrand}, L., \& {Soderblom}, D. 2011, \aj, 142, 23

\bibitem[{{G{\'a}lvez-Ortiz} {et~al.}(2010){G{\'a}lvez-Ortiz}, {Clarke},
  {Pinfield}, {Jenkins}, {Folkes}, {P{\'e}rez}, {Day-Jones}, {Burningham},
  {Jones}, {Barnes}, \& {Pokorny}}]{Galvez2010}
{G{\'a}lvez-Ortiz}, M.~C., {Clarke}, J.~R.~A., {Pinfield}, D.~J., {et~al.}
  2010, \mnras, 409, 552

\bibitem[{{Gray} {et~al.}(2006){Gray}, {Corbally}, {Garrison}, {McFadden},
  {Bubar}, {McGahee}, {O'Donoghue}, \& {Knox}}]{gray2006}
{Gray}, R.~O., {Corbally}, C.~J., {Garrison}, R.~F., {et~al.} 2006, \aj, 132,
  161

\bibitem[{{Gray} {et~al.}(2003){Gray}, {Corbally}, {Garrison}, {McFadden}, \&
  {Robinson}}]{gray2003}
{Gray}, R.~O., {Corbally}, C.~J., {Garrison}, R.~F., {McFadden}, M.~T., \&
  {Robinson}, P.~E. 2003, \aj, 126, 2048

\bibitem[{{Henry} {et~al.}(1996){Henry}, {Soderblom}, {Donahue}, \&
  {Baliunas}}]{henry1996}
{Henry}, T.~J., {Soderblom}, D.~R., {Donahue}, R.~A., \& {Baliunas}, S.~L.
  1996, \aj, 111, 439

\bibitem[{{Jenkins} {et~al.}(2008){Jenkins}, {Jones}, {Pavlenko}, {Pinfield},
  {Barnes}, \& {Lyubchik}}]{jenkins08}
{Jenkins}, J.~S., {Jones}, H.~R.~A., {Pavlenko}, Y., {et~al.} 2008, \aap, 485,
  571

\bibitem[{{Jenkins} {et~al.}(2006){Jenkins}, {Jones}, {Tinney}, {Butler},
  {McCarthy}, {Marcy}, {Pinfield}, {Carter}, \& {Penny}}]{jenkins06}
{Jenkins}, J.~S., {Jones}, H.~R.~A., {Tinney}, C.~G., {et~al.} 2006, \mnras,
  372, 163

\bibitem[{{Jenkins} {et~al.}(2011){Jenkins}, {Murgas}, {Rojo}, {Jones},
  {Day-Jones}, {Jones}, {Clarke}, {Ruiz}, \& {Pinfield}}]{jenkins11}
{Jenkins}, J.~S., {Murgas}, F., {Rojo}, P., {et~al.} 2011, \aap, 531, A8+

\bibitem[{{Jenkins} {et~al.}(2009){Jenkins}, {Ramsey}, {Jones}, {Pavlenko},
  {Gallardo}, {Barnes}, \& {Pinfield}}]{jenkins09}
{Jenkins}, J.~S., {Ramsey}, L.~W., {Jones}, H.~R.~A., {et~al.} 2009, \apj, 704,
  975

\bibitem[{{Khodjachikh} \& {Romanovsky}(2000)}]{Khodjachikh_2000}
{Khodjachikh}, M.~F. \& {Romanovsky}, E.~A. 2000, Kinematika i Fizika Nebesnykh
  Tel, 16, 20

\bibitem[{{L{\'o}pez-Santiago} {et~al.}(2009){L{\'o}pez-Santiago}, {Micela}, \&
  {Montes}}]{lopezsantiago_2009}
{L{\'o}pez-Santiago}, J., {Micela}, G., \& {Montes}, D. 2009, \aap, 499, 129

\bibitem[{{L{\'o}pez-Santiago} {et~al.}(2006){L{\'o}pez-Santiago}, {Montes},
  {Crespo-Chac{\'o}n}, \& {Fern{\'a}ndez-Figueroa}}]{lopezsantiago_2006}
{L{\'o}pez-Santiago}, J., {Montes}, D., {Crespo-Chac{\'o}n}, I., \&
  {Fern{\'a}ndez-Figueroa}, M.~J. 2006, \apj, 643, 1160

\bibitem[{{Makarov} \& {Urban}(2000)}]{Makarov_2000}
{Makarov}, V.~V. \& {Urban}, S. 2000, \mnras, 317, 289

\bibitem[{{Maldonado} {et~al.}(2010){Maldonado}, {Mart{\'{\i}}nez-Arn{\'a}iz},
  {Eiroa}, {Montes}, \& {Montesinos}}]{Maldonado_2010}
{Maldonado}, J., {Mart{\'{\i}}nez-Arn{\'a}iz}, R.~M., {Eiroa}, C., {Montes},
  D., \& {Montesinos}, B. 2010, \aap, 521, A12

\bibitem[{{Mamajek} \& {Hillenbrand}(2008)}]{mamajek&hillenbrand2008}
{Mamajek}, E.~E. \& {Hillenbrand}, L.~A. 2008, \apj, 687, 1264

\bibitem[{{Martin} {et~al.}(2005){Martin}, {Fanson}, {Schiminovich},
  {Morrissey}, {Friedman}, {Barlow}, {Conrow}, {Grange}, {Jelinsky},
  {Milliard}, {Siegmund}, {Bianchi}, {Byun}, {Donas}, {Forster}, {Heckman},
  {Lee}, {Madore}, {Malina}, {Neff}, {Rich}, {Small}, {Surber}, {Szalay},
  {Welsh}, \& {Wyder}}]{galex}
{Martin}, D.~C., {Fanson}, J., {Schiminovich}, D., {et~al.} 2005, \apjl, 619,
  L1

\bibitem[{{Montes} {et~al.}(2001){Montes}, {L{\'o}pez-Santiago}, {G{\'a}lvez},
  {Fern{\'a}ndez-Figueroa}, {De Castro}, \& {Cornide}}]{montes2001}
{Montes}, D., {L{\'o}pez-Santiago}, J., {G{\'a}lvez}, M.~C., {et~al.} 2001,
  \mnras, 328, 45

\bibitem[{{Nakajima} \& {Morino}(2012)}]{Nakajima_2012}
{Nakajima}, T. \& {Morino}, J.-I. 2012, \aj, 143, 2

\bibitem[{{Nordstr{\"o}m} {et~al.}(2004){Nordstr{\"o}m}, {Mayor}, {Andersen},
  {Holmberg}, {Pont}, {J{\o}rgensen}, {Olsen}, {Udry}, \& {Mowlavi}}]{GCSI}
{Nordstr{\"o}m}, B., {Mayor}, M., {Andersen}, J., {et~al.} 2004, \aap, 418, 989

\bibitem[{{Noyes} {et~al.}(1984){Noyes}, {Hartmann}, {Baliunas}, {Duncan}, \&
  {Vaughan}}]{noyes1984}
{Noyes}, R.~W., {Hartmann}, L.~W., {Baliunas}, S.~L., {Duncan}, D.~K., \&
  {Vaughan}, A.~H. 1984, \apj, 279, 763

\bibitem[{{Nutzman} \& {Charbonneau}(2008)}]{nutzman_2008}
{Nutzman}, P. \& {Charbonneau}, D. 2008, \pasp, 120, 317

\bibitem[{{Perryman} {et~al.}(1997){Perryman}, {Lindegren}, {Kovalevsky},
  {Hoeg}, {Bastian}, {Bernacca}, {Cr{\'e}z{\'e}}, {Donati}, {Grenon}, {van
  Leeuwen}, {van der Marel}, {Mignard}, {Murray}, {Le Poole}, {Schrijver},
  {Turon}, {Arenou}, {Froeschl{\'e}}, \& {Petersen}}]{hipparcos}
{Perryman}, M.~A.~C., {Lindegren}, L., {Kovalevsky}, J., {et~al.} 1997, \aap,
  323, L49

\bibitem[{{Pomp{\'e}ia} {et~al.}(2011){Pomp{\'e}ia}, {Masseron}, {Famaey}, {van
  Eck}, {Jorissen}, {Minchev}, {Siebert}, {Sneden}, {L{\'e}pine}, {Siopis},
  {Gentile}, {Dermine}, {Pasquato}, {van Winckel}, {Waelkens}, {Raskin},
  {Prins}, {Pessemier}, {Hensberge}, {Fr{\'e}mat}, {Dumortier}, \&
  {Bienaym{\'e}}}]{Pompeia_2011}
{Pomp{\'e}ia}, L., {Masseron}, T., {Famaey}, B., {et~al.} 2011, \mnras, 415,
  1138

\bibitem[{{Ramsey} {et~al.}(2008){Ramsey}, {Barnes}, {Redman}, {Jones},
  {Wolszczan}, {Bongiorno}, {Engel}, \& {Jenkins}}]{ramsey_2008}
{Ramsey}, L.~W., {Barnes}, J., {Redman}, S.~L., {et~al.} 2008, \pasp, 120, 887

\bibitem[{{Soubiran} \& {Girard}(2005)}]{Soubiran_2005}
{Soubiran}, C. \& {Girard}, P. 2005, \aap, 438, 139

\bibitem[{{Torres} {et~al.}(2006){Torres}, {Quast}, {da Silva}, {de La Reza},
  {Melo}, \& {Sterzik}}]{Torres_2006}
{Torres}, C.~A.~O., {Quast}, G.~R., {da Silva}, L., {et~al.} 2006, \aap, 460,
  695

\bibitem[{{van Leeuwen}(2007)}]{Hippnew}
{van Leeuwen}, F. 2007, \aap, 474, 653

\bibitem[{{Vaughan} \& {Preston}(1980)}]{vaughan80}
{Vaughan}, A.~H. \& {Preston}, G.~W. 1980, \pasp, 92, 385

\bibitem[{{Wright} {et~al.}(2004){Wright}, {Marcy}, {Butler}, \&
  {Vogt}}]{wright2004}
{Wright}, J.~T., {Marcy}, G.~W., {Butler}, R.~P., \& {Vogt}, S.~S. 2004, \apjs,
  152, 261

\bibitem[{{Zuckerman} {et~al.}(2006){Zuckerman}, {Bessell}, {Song}, \&
  {Kim}}]{Zuckerman_2006}
{Zuckerman}, B., {Bessell}, M.~S., {Song}, I., \& {Kim}, S. 2006, \apjl, 649,
  L115

\bibitem[{{Zuckerman} \& {Song}(2004)}]{zuckerman&song_2004}
{Zuckerman}, B. \& {Song}, I. 2004, \araa, 42, 685

\end{thebibliography}


\longtab{4}{
\begin{longtable}{l c c c c c c }
\caption{\label{tblcandidates} Potential moving group members.}\\
\hline\hline
 HD & RA (J2000.0) & DEC (J2000.0) & U [km/s] & V[km/s] & W[km/s]  & $\log R'_{\rm{HK}}$ \\
\hline
\endfirsthead
\caption{continued.}\\
\hline\hline
HD & RA (J2000.0) & DEC (J2000.0) & U [km/s] & V[km/s] & W[km/s]  & $\log R'_{\rm{HK}}$\\
\hline
\endhead
\hline
\endfoot
\hline 
\multicolumn{7}{c}{Pleiades} \\
\hline 
HD 166$^{(1),(2),(4)}$  &  00 06 36.7  &  +29 01 17.4  &  -15.0  &  -22.0  &  -10.0  &  -4.30 \\
HD 17925$^{(2),(3)}$  &  02 52 32.1   &  -12 46 10.9  &  -15.0  &  -21.0  &  -9.0  &  -4.20 \\
HD 26864  &  04 12 43.7  &   -47 33 56.5  &  -14.4  &  -23.0  &  -13.9  &  -4.47 \\
HD 37394$^{(1),(2)}$  &  05 41 20.3  &  +53 28 51.8  &  -13.0  &  -23.0  &  -14.0  &  -4.40 \\
HD 38397  &  05 43 35.8   &  -39 55 24.7  &  -11.0  &  -21.0  &  -6.0  &  -4.26 \\
HD 48286  &  06 42 05.2  &  -15 12 54.9  &  -14.0  &  -22.0  &  -34.0  &  -4.62 \\
HD 100563$^{(1)}$  &  11 34 21.9   &  +03 03 36.5  &  -14.0  &  -21.0  &  -8.0  &  -4.52 \\
HD 116386$^{(4)}$  &  13 24 03.7  &  -45 14 14.2  &  -13.5  &  -23.6  &  -1.0  &  -4.40 \\
HD 130307  &  14 47 16.1  &  +02 42 11.6  &  -15.0  &  -23.0  &  6.0  &  -4.52 \\
HD 202575$^{(1),(3),(4)}$  &  21 16 32.4   &  +09 23 37.7  &  -11.0  &  -20.0  &  -4.0  &  -4.37 \\
HD 206860$^{(1),(2),(4)}$  &  21 44 31.3   &  +14 46 18.9  &  -14.6  &  -21.5  &  -10.8  &  -4.42 \\
HD 220117  &  23 20 53.2   &  +38 10 56.3  &  -15.0  &  -21.0  &  -16.0  &  -4.51 \\

\hline 
\multicolumn{7}{c}{Hyades} \\
\hline 
HD 3196      &  00 35 14.8   &  -03 35 34.2  &  -36.0  &  -19.0  &  -13.0  &  -4.31 \\
HD 5133$^{(3)}$      &  00 53 01.1   &  -30 21 24.9  &  -36.0  &  -20.0  &  13.0  &  -4.60 \\
HD 28406$^{(5)}$    &  04 29 30.3   &  +17 51 47.3  &  -40.0  &  -19.0  &  -1.0  &  -4.46 \\
HD 40979    &  06 04 29.9   &  +44 15 37.5  &  -37.0  &  -21.0  &  8.0  &  -4.63 \\
HD 61033    &  07 34 28.0   &  -52 58 05.3  &  -38.1  &  -22.6  &  4.9  &  -4.42 \\
HD 62911    &  07 44 47.1   &  -33 43 54.3  &  -37.0  &  -22.0  &  -3.0  &  -4.47 \\
HD 73256    &  08 36 23.0   &  -30 02 15.4  &  -36.0  &  -21.0  &  -15.0  &  -4.52 \\
HD 87978$^{(4)}$    &  10 08 26.5   &  -11 06 54.7  &  -37.6  &  -23.2  &  0.9  &  -4.45 \\
HD 92855    &  10 44 00.6   &  +46 12 23.9  &  -40.0  &  -21.0  &  -14.0  &  -4.44 \\
HD 106489$^{(6)}$  &  12 14 57.4   &  -41 08 21.9  &  -38.0  &  -23.0  &  -18.0  &  -4.48 \\
HD 111998  &  12 53 11.1   &  -03 33 11.1  &  -36.0  &  -20.0  &  -6.0  &  -4.43 \\
HD 132301  &  14 59 45.0   &  -43 48 40.8  &  -39.0  &  -23.0  &  -8.0  &  -4.35 \\
HD 140011  &  15 43 11.8   &  -44 43 12.3  &  -38.0  &  -19.0  &  -7.0  &  -4.46 \\
HD 153330$^{(6)}$  &  17 00 25.4   &  -37 56 14.0  &  -36.0  &  -19.0  &  19.0  &  -4.64 \\
HD 215532  &  22 47 06.7   &  -60 04 33.2  &  -40.0  &  -22.0  &  -4.0  &  -4.68 \\

\hline 
\multicolumn{7}{c}{Sirius/UMa} \\
\hline 
HD 25926$^{(6)}$    &  04 04 55.9   &  -35 26 47.6  &  8.0  &  3.0  &  -10.0  &  -4.45 \\
HD 30059    &  04 43 47.7   &  -12 17 40.0  &  5.2  &  1.6  &  1.0  &  -4.60 \\
HD 152388  &  16 56 52.1   &  -60 57 24.6  &  8.0  &  5.0  &  4.0  &  -4.62 \\ 
HD 187532  &  19 50 46.7   &  -10 45 48.6  &  5.0  &  5.0  &  3.0  &  -4.09 \\ 
HD 215657$^{(6)}$  &  22 47 26.7   &  -44 57 54.5  &  9.0  &  2.0  &  -7.0  &  -4.28 \\

\hline 
\multicolumn{7}{c}{Coma Berenices + Castor} \\
\hline 
HD 8907    &  01 28 34.3  &  +42 16 03.6  &  -10.0  &  -4.0  &  -17.0  &  -4.50 \\
HD 10611$^{(6)}$  &  01 43 14.3  &  -21 37 11.1  &  -7.0  &  -7.0  &  -8.0  &  -4.30 \\
HD 13507$^{(1),(4)}$  &  02 12 55.0  &  +40 40 06.0  &  -9.0  &  -6.0  &  -11.0  &  -4.39 \\
HD 13531  &  02 13 13.3  &  +40 30 27.3  &  -10.0  &  -6.0  &  -11.0  &  -4.34 \\ 
HD 18809$^{(6)}$  &  03 00 19.7  &  -37 27 16.1  &  -9.0  &  -4.0  &  -10.0  &  -4.33 \\
HD 21175  &  03 23 35.2  &  -40 04 34.9  &  -8.0  &  -7.0  &  -10.0  &  -4.62 \\
HD 21722$^{(7)}$  &  03 25 36.2  &  -69 20 11.1  &  -7.0  &  -7.0  &  -15.0  &  -4.31 \\
HD 25120  &  03 55 18.3  &  -65 17 05.0  &  -9.0  &  -5.0  &  -13.0  &  -4.40 \\
HD 33256$^{(8)}$  &  05 08 43.7  &  -04 27 22.3  &  -10.0  &  -6.0  &  2.0  &  -4.61 \\
HD 103978  &  11 58 19.5  &  -55 14 22.4  &  -8.4  &  -5.1  &  -8.5  &  -4.46 \\

\hline
\multicolumn{7}{c}{NGC 1901} \\
\hline
HD 7678      &  01 15 45.0  &  -53 42 56.4  &  -25.0  &  -11.0  &  1.0  &  -4.70 \\
HD 10655      &  01 43 32.3  &  -19 24 12.6  &  -24.4  &  -9.4  &  -6.41 &  -4.44 \\
HD 30157    &  04 44 33.8  &  -16 03 54.1  &  -21.6  &  -11.1  &  -3.9  &  -4.67 \\
HD 43162$^{(1),(3),(4)}$    &  06 13 45.2  &  -23 51 42.9  &  -21.0  &  -10.0  &  -7.0  &  -4.33 \\
HD 57852$^{(9)}$    &  07 20 21.5  &  -52 18 43.2  &  -22.0  &  -13.0  &  -8.0  &  -4.16 \\
HD 84273    &  09 43 20.2  &  -29 48 14.4  &  -22.0  &  -10.0  &  -5.0  &  -4.50 \\
HD 118036  &  13 34 16.2  &  -00 18 49.6  &  -23.0  &  -12.0  &  0.0  &  -4.57 \\
HD 142415$^{(6)}$  &  15 57 40.7  &  -60 12 00.9  &  -24.0  &  -13.0  &  1.0  &  -4.62 \\
HD 162396$^{(8)}$  &  17 52 52.5  &  -41 59 47.4  &  -24.0  &  -11.0  &  -32.0  &  -4.64 \\
HD 218168  &  23 04 02.1  &  +74 28 38.7 &  -25.0  &  -9.0  &  -2.0  &  -4.50 \\ 

\hline
\multicolumn{7}{c}{IC 2391} \\
\hline

HD 25680$^{(1),(3)}$    &  04 05 20.2  &  +22 00 32.0  &  -25.0  &  -14.0  &  -6.0  &  -4.45 \\
HD 53143$^{(1),(4)}$    &  06 59 59.6  &  -61 20 10.2  &  -25.0  &  -18.0  &  -15.0  &  -4.39 \\
HD 61005    &  07 35 47.4  &  -32 12 14.0  &  -22.0  &  -14.0  &  -4.0  &  -4.18 \\
HD 62850$^{(6),(10)}$    &  07 42 36.0  &  -59 17 50.7  &  -25.0  &  -18.0  &  -2.0  &  -4.27 \\
HD 62848$^{(9),(10)}$    &  07 43 21.4  &  -52 09 50.7  &  -25.0  &  -18.0  &  -3.0  &  -4.33 \\
HD 103742$^{(4),(10)}$  &  11 56 42.3  &  -32 16 05.4  &  -24.0  &  -18.0  &  -4.0  &  -4.33 \\
HD 110419  &  12 42 08.8  &  -31 49 10.6  &  -21.1  &  -18.1  &  -16.2  &  -4.48 \\
HD 113553$^{(6)}$  &  13 05 16.8  &  -50 51 23.8  &  -23.0  &  -17.0  &  2.0  &  -4.30 \\
HD 120352$^{(4)}$  &  13 48 58.1  &  -01 35 34.6  &  -21.2  &  -18.5  &  -10.0  &  -4.64 \\
HD 151598  &  16 50 46.8  &  -49 12 39.3  &  -25.0  &  -15.0  &  -8.0  &  -4.40 \\
HD 158866  &  17 47 42.6  &  -82 12 59.1  &  -25.0  &  -17.0  &  -4.0  &  -4.35 \\
HD 200361  &  21 04 31.2  &  -44 27 42.4  &  -22.4  &  -17.9  &  -11.8  &  -4.59 \\
HD 203030$^{(4)}$  &  21 18 58.2  &  +26 13 49.9  &  -23.0  &  -16.0  &  -12.0  &  -4.37 \\

\hline
\multicolumn{7}{c}{Dehnen 7} \\
\hline
HD 8326      &  01 22 07.6  &  -26 53 35.1  &  24.0  &  -23.0  &  -22.0  &  -4.65  \\
HD 21899    &  03 30 13.5  &  -41 22 11.7  &  20.0  &  -23.0  &  -13.0  &  -4.60 \\
HD 137778$^{(3)}$  &  15 28 12.2  &  -09 21 28.2  &  22.0  &  -22.0  &  -20.0  &  -4.32 \\

\hline
\multicolumn{7}{c}{Dehnen 9} \\
\hline
HD 11683  &  01 54 22.1  &  -15 43 26.3  &  -24.1  &  -51.5  &  8.8  &  -4.46 \\
HD 73744  &  08 32 15.5  &  -76 55 43.6  &  -22.0  &  -50.0  &  -31.0  &  -4.68 \\

\hline
\multicolumn{7}{c}{Dehnen 10} \\
\hline
HD 64184    &  07 49 26.6  &  -59 22 51.0  &  52.0  &  -2.0  &  28.0  &  -4.63 \\
HD 71251    &  08 26 11.9  &  -05 45 05.4  &  50.3  &  -0.2  &  -6.1  &  -4.58 \\
HD 192117  &  20 20 24.1  &  -75 34 40.3  &  52.0  &  0.0  &  4.0  &  -4.67 \\

\hline
\multicolumn{7}{c}{Dehnen 11} \\
\hline
HD 37962  &  05 40 51.9  &  -31 21 03.9  &  54.0  &  -24.0  &  -26.0  &  -4.627 \\ 

\hline
\multicolumn{7}{c}{$\zeta$ Herculis} \\
\hline
HD 59984$^{(8)}$    &  07 32 05.7  &  -08 52 52.7  &  -29.0  &  -51.0  &  -18.0  &  -4.51 \\

\hline

\end{longtable}
\tablefoot{(1)~\citet{Nakajima_2012}; (2) \citet{lopezsantiago_2006}; (3) \citet{Maldonado_2010}; (4) \citet{montes2001}; (5) \citet{boesgaard1988}; (6) \citet{Torres_2006}; (7) \citet{Khodjachikh_2000}; (8) \citet{Soubiran_2005}; (9) \citet{Makarov_2000}; (10) \citet{Zuckerman_2006}.}
}

\end{document}